\documentclass[iop]{emulateapj}

\usepackage{epsf}    
\usepackage{graphicx}
\usepackage{epstopdf}

\usepackage{color}
\usepackage{amsmath}
\usepackage{graphicx}
\usepackage{amssymb}
\usepackage{psfrag}

\usepackage{natbib,times}
\newcommand{\beq}{\begin{equation}}
\newcommand{\eeq}{\end{equation}}
\newcommand{\be}{{\bf{e}}}
\renewcommand{\bv}{{\bf{v}}}
\newcommand{\nn}{\nonumber}
\newcommand{\bB}{{\bf{B}}}

\newcommand{\rmd}{\mathrm{d}}
\newcommand{\brac}[1]{\left({#1}\right)}
\newcommand{\pd}[2]{\frac{\partial{#1}}{\partial{#2}}}
\newcommand{\td}[2]{\frac{\rmd{#1}}{\rmd{#2}}}
\newcommand{\curl}{\nabla\times}
\renewcommand{\div}{\nabla\cdot}

\newcommand{\bpsi}{\boldsymbol{\psi}}

\shorttitle{Magnetar field evolution and crustal plasticity}
\shortauthors{Lander}

\begin{document}

\title{Magnetar field evolution and crustal plasticity}

\author{S. K. Lander}
\email{skl@soton.ac.uk}
\affil{Mathematical Sciences and STAG Research Centre, University of
  Southampton, Southampton SO17 1BJ, UK}

\begin{abstract}
The activity of magnetars is believed to be powered by
colossal magnetic energy reservoirs. We sketch an evolutionary picture in which
internal field evolution in magnetars generates a twisted corona,
  from which energy may be released suddenly in a single giant flare,
  or more gradually through smaller outbursts and persistent
  emission. Given the ages of magnetars and the energy of their giant
flares, we suggest that their evolution is driven by a novel
mechanism: magnetic flux transport/decay due to
persistent plastic flow in the crust, which would invalidate the common
assumption that the crustal lattice is static and evolves only under
Hall drift and Ohmic decay. We estimate the field strength required to
induce plastic flow as a function of crustal depth, and the
viscosity of the plastic phase. The star's superconducting
core may also play a role in magnetar field evolution, depending on
the star's spindown history and how rotational vortices and magnetic
fluxtubes interact.
\end{abstract}

\keywords{dense matter --- magnetic fields --- stars: flare --- stars: magnetars --- stars: neutron}

\maketitle


\section{Introduction}

Some of the most
intriguing neutron stars are the \emph{magnetars}: highly magnetised
objects whose surface fields are inferred to be in
excess of $10^{14}$ G in some cases, and whose interior fields may
reach $10^{16}$ G \citep{TD95,turolla-review}.
In contrast with many older, more predictable neutron stars,
magnetars are volatile, alternating between quiescent states and highly energetic
bursts and flares. Their most spectacular events are the giant flares,
releasing over $\sim 10^{45}$ erg of energy in a very brief flash and
decaying X-ray tail. Three of
these frustratingly rare events have been seen to date, from three
different magnetars -- giving little idea of their event rate.

As with much of magnetar physics, there is a broadly-accepted qualitative
explanation for giant flares, but filling in the details
is challenging. Storing and releasing the requisite amount of energy
is believed to begin with
internal magnetic field evolution building stresses in the
magnetar's crust. Here we try to produce a
quantitative description of this, and are led to one of two possible
scenarios for the star's magnetic-field evolution: it is either driven
by persistent plastic flow in the crust, or rotational vortices
dragging out superconducting fluxtubes from the core as the star spins down.

\subsection{From flare to corona to interior}

The initial rise timescale for giant flares is 1ms,
suggestive of some explosive reconnection process in the charge-filled
corona surrounding the 
star \citep{lyu03,elenbaas,huang_yu}, in analogy with solar flares
\citep{masada}. For the timescale to be sufficiently short, this
reconnection must occur within a few stellar radii ($\sim
10$km) of the surface. Direct release of energy from the crust is
unlikely: the shortest characteristic timescale is $\sim 0.2$s, from
shear-wave propagation -- and the release of energy
may be slower still \citep{lyu06,link14}.

The coronal-reconnection scenario for giant flares requires a huge
amount of magnetic energy to be stored in strongly-twisted exterior field lines.
The long periods of magnetars, $\sim 1-10$s, mean that the
rotation-powered mechanism invoked for pulsar magnetospheres would
only twist a small region of field lines near the polar cap \citep{GLA}. Instead,
shearing of the crust could move magnetospheric 
footpoints \citep{TLK02}, inducing a twist and
forming an equatorial lobe of current,
with charges stripped off the surface \citep{belo09}.

This corona, in turn, requires a mechanism for build-up of crustal
stresses. If the star were born rotating rapidly, its crust would freeze into an
oblate spheroid shape, and the star's later spindown would induce
stresses as the preferred shape of the star became more spherical
\citep{rud69} -- but this axisymmetric process would displace
footpoints vertically, without inducing any coronal twist. The only
credible candidate to generate the required azimuthal crustal motion is
the star's magnetic field. Magnetar-corona simulations
show how crustal motion (albeit added by hand) produces
exterior twist \citep{parfrey}, potentially 
leading to an overtwisting instability \citep{wolfson} and flare. The build-up of
stresses due to a changing global equilibrium can produce
the right kind of crustal displacement \citep{LAAW}, and
power giant flares, but this has not been 
verified with full core-crust field evolutions. On the other hand,
magneto-thermal evolutions of the crust show how the crustal field can
evolve to become locally intense and fail in small events
\citep{ponsrea}. Whilst this provides a credible explanation of the
phenomena of magnetar bursts and their persistent luminosity, it is
less clear whether it can be applied to giant flares, with their
shorter rise timescales and greater energy release.

\subsection{Energy budget and storage}

The relevant region for energy
storage is whatever portion of the interior stellar magnetic
field can be tapped for giant flare energy; a deeply-buried core field
may be irrelevant over magnetar lifetimes.
The most energetic giant flare, from SGR 1806-20, was
estimated to be $3\times 10^{46}$ erg \citep{hurley}; the other two giant
flares, from SGR 1900+14 \citep{feroci} and SGR 0526-66
\citep{mazets}, were both $\sim 10^{45}$
erg. In order to have a mechanism 
that can credibly provide enough energy to power any giant flare, we
will adopt a magnetar model harbouring ten times the energy of the
largest one observed: $3\times 10^{47}$ erg. Note that the more detailed
  discussion in \citet{TD01} suggests a magnetar energy budget
  $\gtrsim 10^{47}$ erg; this is probably a conservative value, since
  the paper predated the largest giant flare.


\section{Crust-only field evolution implies plastic flow}

Evolution of the field in a neutron star (NS) core is not fully
understood, and may not be fast enough to explain magnetar
activity (see section 3). It is thus natural to begin by 
examining crustal evolution, and to assume that the
core does not contribute to the energy powering giant flares.
A magnetic field $\bB$ confined to the crust, with the $3\times 10^{47}$ erg
of energy we require, must have the following
average strength $\langle B\rangle$:
\beq \label{crust_field}
3\times 10^{47}\textrm{ erg}\lesssim \int \frac{B^2}{8\pi}\rmd V
 \implies\langle B\rangle\gtrsim 3\times 10^{15}\textrm{ G}.
\eeq
Since models of crustal fields harbour a maximum $B$ which is a few
times $\langle B\rangle$ \citep{gour13}, we may expect $B\sim 10^{16}$ G locally.
If we had instead allowed the field to extend throughout the star,
$\langle B\rangle$ would (roughly) halve. Although 
dipole-field estimates from spindown can be unreliable 
\citep{younes}, this suggests that values around $10^{15}$ G are credible.

A field confined to the crust will not satisfy global hydromagnetic
equilibrium and must be supported by
elastic stresses. Indeed, the estimate above is for a field which
\emph{is} stressing the crust, since we eventually require this
magnetic energy to be released into the corona. However, such a strong
field will stress the crust beyond its elastic limit
\citep{LAAW,pernapons,TD95}. Therefore, if we assume the core's energy does not contribute
to giant flares, then the crust must undergo magnetically-induced
plastic flow \citep{jones06}.

\subsection{Equations of motion for magnetically-driven plastic flow}

We assume that a NS crust responds in an elastic, reversible manner to stresses below
some yield value $\tau_\textrm{el}$ and plastically above it. Many terrestrial media can also be approximated as having
elastic or plastic responses, depending on the magnitude of the applied stress; they
include toothpaste, crude oil, mud and concrete.

Let us first consider a magnetised crust below $\tau_\textrm{el}$. $\bB$ need
not be in a global fluid equilibrium 
with the star, since the crust can absorb magnetic stresses:
\beq \label{eom_eqm}
0 = -\nabla P-\rho\nabla\Phi+\frac{1}{4\pi}(\curl\bB)\times\bB+\div\boldsymbol{\tau},
\eeq
where $P$ is pressure, $\rho$ mass density, $\Phi$ the gravitational potential and
$\boldsymbol\tau$ the elastic stress tensor. We will think of the
crust reaching $\tau_\textrm{el}$ for some Lorentz force
  $(\curl\bB_\textrm{el})\times\bB_\textrm{el}$, which defines the
`yield magnetic field' $\bB_\textrm{el}$. For $\bB>\bB_\textrm{el}$ the crust's response to the force
\beq \label{DeltaB}
(\curl\Delta\bB)\times\Delta\bB\equiv 
 (\curl\bB)\times\bB-(\curl\bB_\textrm{el})\times\bB_\textrm{el}
 > 0
\eeq
will be plastic, unless the crust is in hydromagnetic equilibrium (for
which $\boldsymbol{\tau=0}$).  Equation \eqref{DeltaB} defines the
field $\Delta\bB$ which sources the plastic flow; note that in general
$\Delta\bB\neq\bB-\bB_\textrm{el}$.

\begin{figure}
\psfrag{Bel}{$\displaystyle{\frac{B_\textrm{el}}{10^{15}\textrm{ G}}}$}
\psfrag{r}{$\displaystyle{r/R_*}$}
\centerline{\includegraphics[width=0.4\textwidth]{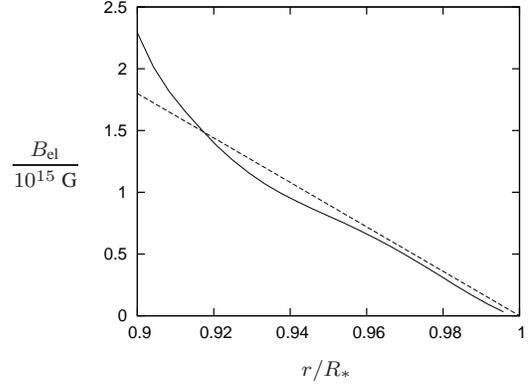}}
\caption{Solid line: the radial dependence of the field strength $B_\textrm{el}$
  required to induce plastic flow in the crust, as defined in equation
  \eqref{Bel_defn}. Dashed line: an approximation to this quantity,
  equation \eqref{Bel_approx}, where $B_\textrm{el}$
  varies linearly with depth.
\label{Bel}}
\end{figure}

Next we wish to make a depth-dependent estimate of
$B_\textrm{el}$, rather than using the standard assumption that
$B_\textrm{el}\sim 10^{15}$ G. As in
\citet{LAAW}, we make fits to NS equation-of-state parameters 
from \citet{douchin} and a magnetar temperature profile from
\citet{kaminker}.
Using these, we calculate $\tau_\textrm{el}$ from the formula of 
\citet{chughoro}. Now, equation \eqref{eom_eqm} evaluated at the yield
stress shows us that a natural definition for $B_\textrm{el}$ is:
\beq \label{Bel_defn}
B_\textrm{el}\equiv \sqrt{4\pi\tau_\textrm{el}}.
\eeq
Plotting this in figure \ref{Bel}, we see that $B_\textrm{el}$ may be
approximated by the linear relation
\beq \label{Bel_approx}
B_\textrm{el} \approx 1.8\times 10^{16}\brac{1-\frac{r}{R_*}}\textrm{ G},
\eeq
where $r$ is the radial coordinate and $R_*$ the surface radius.

Beyond the elastic yield stress, the crust must be in motion, with
some velocity $\mathbf{v}$. Since we expect this
flow to be approximately incompressible, we take
$\div\mathbf{v}=0$. 
General models of viscoplastic flow are based on a relation between two
tensorial quantities: the rate of strain
$\boldsymbol{\dot\varepsilon}$ and the stress.
Let us however assume that the crust is under simple shear
  stress, for which each tensor has a single independent component, $\dot{\varepsilon}$
  and $\tau$, and the problem reduces to a scalar one. We may now adopt
  the Bingham model to relate $\dot\varepsilon$ and $\tau$ when $\tau>\tau_\textrm{el}$:
\beq \label{bingham_scal}
\dot\varepsilon=\frac{1}{2\nu}(\tau-\tau_\textrm{el})H(\tau-\tau_\textrm{el}),
\eeq
where $\nu$ is the dynamical viscosity of the plastic flow and $H(\cdot)$ the Heaviside function. 
Under equation \eqref{bingham_scal}, one can derive an equation of
motion valid above $\tau_\textrm{el}$ \citep{prager}; for our magnetar crust model
this is:
\begin{align}
\rho\dot\bv &+ \rho(\bv\cdot\nabla)\bv \nn\\
 &= -\nabla P - \rho\nabla\Phi+\nu\nabla^2\bv
        + \frac{1}{4\pi}(\curl\bB)\times\bB +\div\boldsymbol{\tau}_\textrm{el}.
 \label{eom_plastic}
\end{align}
The fluid terms in this equation have the same form as in the standard
Navier-Stokes equation for a viscous medium. In general, however,
viscoplastic dynamics can be richer than those of viscous fluids --
if, for example, the medium yields in a more complicated manner than
through simple shearing, a tensor generalisation
of equation \eqref{bingham_scal} is required. This in turn results in
the appearance of a new, uniquely plastic force term in the equation of motion,
$\tau_\textrm{el}\div(\dot{\boldsymbol{\varepsilon}}/|\!|\dot{\boldsymbol{\varepsilon}}|\!|)$,
where $|\!|\cdot|\!|$ is a tensor norm \citep{prager}. This may be
important for more sophisticated modelling of NS crustal
failure.

Returning to equation \eqref{eom_plastic}, we anticipate the plastic
flow to be slow 
and steady, and hence neglect the advective term
$(\bv\cdot\nabla)\bv$ for being quadratic in $\bv$
and the acceleration term $\dot\bv=0$. The resulting viscous crustal 
motion is \emph{Stokes flow}. Comparing this limiting case of equation
\eqref{eom_plastic} with \eqref{eom_eqm}, and dropping the tiny
  differences between the hydrostatic terms $\nabla P+\rho\nabla\Phi$
  in the elastic and plastic regimes, we arrive at the intuitive
result that the unbalanced piece of the Lorentz force sources a
viscous flow:
\beq \label{stokes}
\nu\nabla^2\bv = - \frac{1}{4\pi}(\curl\Delta\bB)\times\Delta\bB
\ \ ,\ \ \div\bv=0.
\eeq
If this unbalanced Lorentz force is curl-free, equation \eqref{stokes} takes an extremely compact form. Since
$\div\bv=0$ we may write $\bv=\curl\bpsi$ for some potential
$\bpsi$. This $\bpsi$ is only fixed
up to transformations of the form $\bpsi\to\bpsi+\nabla\phi$, a
freedom which allows us to choose $\div\bpsi=0$ (the Coulomb
gauge). Now taking the curl of equation \eqref{stokes} and using these results,
together with various vector identities (including those for a
triple curl), we find that the flow is governed by the
biharmonic equation:
\beq \label{biharm}
\nabla^4\bpsi=0.
\eeq

\subsection{Timescale for plastic-flow-induced field evolution}

The overwhelming majority of studies 
of magnetised NS crusts assume \emph{electron
  magnetohydrodynamics}: the crustal ion lattice is strictly static, magnetic
forces are balanced by elastic stresses (rendering the
equation of motion irrelevant), and only the electrons are mobile
(see, e.g., \citet{gour,wood}). Field evolution is then governed
by the interplay of two secular terms \citep{cumming}: the
conservative Hall drift, and Ohmic decay
(the second and third terms, respectively, in equation
\eqref{induction}). One situation where this is clearly no longer valid is for
$\tau>\tau_\textrm{el}$, when we need to add a third term involving
the plastic-flow velocity $\bv$:
\beq \label{induction}
\pd{\bB}{t} = \curl(\bv\times\bB)
                      - \curl\left[\frac{(\curl\bB)\times\bB}{4\pi \rho_e}\right]
                      - \curl\left[\frac{\curl\bB}{4\pi\sigma_0}\right],
\eeq
where $\rho_e$ is charge density and $\sigma_0$ electrical conductivity.
Next we check when this
plastic-flow term may dominate the crust's evolution.
Dimensional analysis of equation \eqref{stokes} shows that
\beq \label{v_dim}
v \sim \frac{L_\textrm{pl}^2(\Delta B)^2}{L_\textrm{B}\nu}
\eeq
where $L_\textrm{pl},L_\textrm{B}$ are the lengthscales of the plastic flow and
the field, respectively. Assuming the plastic flow to be more
localised than the magnetic field, i.e. $L_\textrm{pl}\leq
L_\textrm{B}$, and also that $\Delta B\sim B$
(reasonable unless the field is just above $B_\textrm{el}$),
equations \eqref{v_dim} and \eqref{induction} combine
to give the following plastic-flow evolution timescale:
\beq
t_\textrm{pl} \sim \frac{L_\textrm{B}}{L_\textrm{pl}}\frac{\nu}{B^2},
\ \textrm{ when }B>B_\textrm{el}. 
\eeq
The plastic viscosity $\nu$ is an unknown crustal parameter, whose
calculation might require molecular-dynamics simulations, but we
may make an estimate by arguing from dimensional analysis that 
$\nu\sim t_\textrm{char}\tau_\textrm{el}$ for some characteristic
plastic timescale $t_\textrm{char}$. We already know
$\tau_\textrm{el}$, from the calculation described before
\eqref{Bel_defn}. For $t_\textrm{char}$, let us demand it
be short enough to allow for a persistent corona. Since coronal
dissipation occurs over $\sim 1-10$ yr \citep{BT07}, we need to take
$t_\textrm{char}=10$ yr to allow sufficiently fast
regeneration of the exterior current.
$\nu$ will increase with crustal depth and decrease with
temperature, but rather than exploring these dependences here we will
simply evaluate $t_\textrm{char}\tau_\textrm{el}$ at
the crust-core boundary for our (hot) magnetar model, and use
this as a reference value: $\nu\sim 10^{38}$ poise.
If this estimate is reliable, then the $\nu\lesssim 10^{35}$-poise condition for thermo-plastic instability
\citep{BL14} would only be attained in the outer crust. For 
comparison, a famously viscous terrestrial material is
pitch; in an experiment begun in 1930, nine drops of pitch have fallen to date, yielding
a viscosity estimate of $2\times 10^9$ poise \citep{pitch}.

Using our viscosity estimate, we may compare $t_\textrm{pl}$ with the
timescale $t_\textrm{Hall}$ for Hall drift (which is
faster than Ohmic decay for high $B$), using the results of
\citet{cumming}. We forsee three regimes:\\
(1) if $B\gtrsim 10^{15}$ G, plastic flow
always dominates, with a characteristic timescale $\lesssim 3
L_\textrm{B}/L_\textrm{pl}$ yr;\\
(2) in the range $B\sim (2-7)\times 10^{14}$ G, $B_\textrm{el}$ is
exceeded for crustal depths $\lesssim 100-400$m, and 
$t_\textrm{pl}\sim t_\textrm{Hall}\sim 10-100$ yr there, so the two
field-evolution mechanisms compete. Deeper into the crust there is no
plastic flow;\\
(3) if $B\lesssim 10^{13}$ G, plastic flow shuts off essentially
everywhere.

\subsection{A toy model}

\begin{figure}
\centerline{\includegraphics[width=0.3\textwidth]{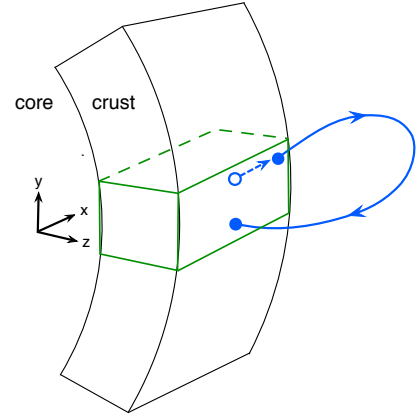}}
\caption{Toy-model geometry. A segment of crust assumed to undergo
  plastic flow is approximated by a slab in Cartesian geometry, with
  axes orientated as shown. Azimuthal motion of the crust
  ($x$-direction flow in the slab) moves one footpoint of an initially
  poloidal coronal field line, twisting it as shown.
\label{toy_geom}}
\end{figure}

We now consider a simple illustrative example of magnetically-induced
plastic flow that can be solved analytically. Taking a segment of the crust
and neglecting its curvature, we are left with a Cartesian slab of
material, as shown in figure \ref{toy_geom}:
the $x$ and $z$ coordinates represent azimuthal and radial
directions in the crust, respectively.

Assume a Lorentz force which stresses the crust beyond
$\tau_\textrm{el}$ in some band $-y_0<y<y_0$. To produce a constant,
non-radial force we choose
\beq
\Delta\bB = b\sqrt{x}\be_z,
\eeq
where $b$ is a constant. Note that $\Delta\bB$ is divergence-free, and its
corresponding force curl-free, as required. Our assumptions, however,
force us to have a depth-independent
$\nu$; the intention is to relax this restriction in future work.
Now, returning to equation \eqref{stokes},
\beq \label{laplace}
\nabla^2\bv = (\nabla^2 v_x)\be_x + (\nabla^2 v_y)\be_y
                     = \frac{b^2}{8\pi\nu}\be_x.
\eeq
Here $v_z=0$ to ensure incompressibility of the flow. Equation
\eqref{laplace} implies $\nabla^2 v_y=0$; we will satisfy this by taking
$v_y=0$ (this is not actually a simplification; one can show that
keeping a $v_y\neq 0$ term is incompatible with confining the flow
into a channel $-y_0<y<y_0$).

In terms of the velocity potential,
$\bv=\be_x\partial\psi/\partial y-\be_y\partial\psi/\partial x$,
so $v_y=0$ implies $\psi=\psi(y)$, and equation \eqref{biharm} becomes simply
$\rmd^4 \psi/\rmd y^4=0$.
Integrating once and comparing with equation \eqref{laplace} gives
\beq
\td{^3 \psi}{y^3} = \td{^2 v_x}{y^2}=\frac{b^2}{8\pi\nu}.
\eeq
Integrating twice more, and imposing the boundary conditions that
the flow must stop when there is no unbalanced Lorentz force,
i.e. $v_x(-y_0)=v_x(y_0)=0$, we find that the plastic-flow velocity is
\beq
\bv = \frac{b^2}{16\pi\nu}(y^2-y_0^2) \be_x.
\eeq
The evolution of a crustal field
$\bB=B_x(x,y,t)\be_x+B_y(x,y,t)\be_y+B_z(x,y,t)\be_z$ under this flow
is given by $\curl(\bv\times\bB)$, i.e.
\begin{align}
\pd{\bB}{t}
  = \frac{b^2}{16\pi\nu}
          \bigg\{ & \pd{ }{y}[(y^2-y_0^2) B_y] \be_x \nn\\
                     & + (y_0^2-y^2) \brac{\pd{B_y}{x}\be_y + \pd{B_z}{x} \be_z}
          \bigg\}.
\end{align}
By separation of variables, writing each field component in the form
$X(x)Y(y)T(t)$, we find that we require $B_y=B_y(y)$ and $B_z=B_z(y)$
for consistency, so that only $B_x$ is time-varying, with
\beq
B_x=\frac{b^2}{16\pi\nu} \td{ }{y}[(y^2-y_0^2) B_y] t.
\eeq
This toy model suggests that a magnetic field with an unbalanced
radial component induces an azimuthal plastic flow, whose velocity
depends (through $b$) on how much one exceeds $\tau_\textrm{el}$; field
evolution along this flow provides an ever-increasing twist to the
corona. More realistically, the twist would be relieved gradually
through viscoplastic or Ohmic dissipation, or suddenly in
coronal flare events.


\section{Core-field evolution driven by stellar spindown}

We may be able to avoid persistent plastic crustal flow if the core
field contributes to the magnetar's usable energy reservoir, so next we
review core-field evolution mechanisms to identify any which may be
sufficiently rapid. We report timescales for a field lengthscale
$L_\textrm{B}=1$km, the crust-core boundary density $1.28\times
10^{14}$ g cm${}^{-3}$, 
and normalise to a temperature $T_8\equiv T/(10^8\textrm{ K})$ and
field strength $B_{15}\equiv B/(10^{15}\textrm{ G})$.

A normal-matter core with protons, neutrons and electrons has
three field-evolution mechanisms \citep{GR92}. Two of these --
Ohmic decay ($\sim 10^{11} T_8^{-2}$yr) and Hall drift ($\sim 5\times
10^5 B_{15}^{-1}$yr) -- are too slow for our purposes. The third, ambipolar diffusion, appears
more promising: its solenoidal component could cause a buoyant rise of flux out of the
core over $\sim 10^3 T_8^2 B_{15}^{-2}$ yr. 
However, the normal-fluid model is not applicable to any known NS; all
have cooled sufficiently to harbour superfluid neutrons and type-II
superconducting protons \citep{HoGA}. Neutron superfluidity
drastically alters the action of ambipolar diffusion: the solenoidal
drift becomes too fast to be relevant, whilst the other drift
component acts over $\sim 10^{20}$ yr \citep{GJS}.
Proton superconductivity also affects field evolution, although
one can derive analogues of Hall drift and Ohmic decay, assuming
the dissipative mechanism is mutual friction \citep{graber}. The situation is
no more promising for rapid field evolution, however, giving a minimum
timescale of $\sim 10^6$ yr. The mostly
unexplored physics of the crust-core boundary may allow for rapid
field evolution, or inhibit it \citep{konenk_gepp}; here we
ignore this important open issue and focus instead on the only
potentially rapid core-evolution mechanism we are aware of.

In the superfluid-superconducting outer NS core, bulk stellar rotation
is quantised into neutron vortices, and the magnetic field into
fluxtubes, provided the magnetic field is below the value $H_{c2}\sim 10^{16}$ G at which
superconductivity is destroyed. Vortices and fluxtubes cannot
generally move far without encountering one another. If the energy
penalty associated with them cutting through each other is
sufficiently large, they will instead `pin' together, and their motion
will be coupled \citep{sauls89,link03,gug_alp}. In particular, as the star spins down
vortices will move outwards, and in turn drag fluxtubes with them \citep{rud98}. This
could lead to core-field evolution over a timescale
as short as $\sim 10^4$yr, though \citet{jones06} and \citet{GA11}
have reached opposing conclusions about whether this mechanism would
operate efficiently in magnetars. Even if it does though, the extent to which
vortex-fluxtube `pinning' can induce magnetic 
stresses depends strongly on the star's spin-down history; rapid core field evolution in a
magnetar would suggest that it had been born rapidly-rotating.


\section{Discussion and implications}

We have argued that magnetar field evolution -- in particular for the
objects who have suffered giant flares -- cannot be entirely driven by
the conventional evolution mechanisms of Ohmic decay and Hall
drift. Instead, if the evolution proceeds mainly in the crust then
magnetic stresses are likely to become large enough to induce
evolution through persistent plastic flow. The role of core-field evolution for
magnetars is less clear: it depends on whether vortices and fluxtubes
`pin' to one another, and relies on the protons forming a
type-II superconductor, which occurs if $B<H_{c2}\sim 10^{16}$ G. This latter
condition represents another uncertainty: not only are magnetar core
fields unknown, but calculations of $H_{c2}$ also vary over an
order of magnitude (see, e.g., \citet{GAS} and \citet{sinsed}).

The two scenarios suggest different mechanisms for generating a magnetar
corona and triggering a giant flare. With core evolution, the crust
could respond
elastically as stresses build, then fail rapidly; in this case, we
anticipate the immediate formation of a transient corona, which may
be dynamically unstable and result in a giant flare. Crustal evolution leads to a more
permanent corona, sourced by shifting magnetospheric footpoints
embedded in the plastically-deforming crust, and a giant flare might represent
an overtwisting instability. The presence of long-lived magnetar coronae
is consistent with the persistently high spindown rate following the giant
flare of SGR 1806-20 \citep{younes}, and the X-ray spectra of a number
of other magnetars \citep{weng}. 
Note that \citet{lyu15} also briefly discusses plastic-flow-driven
field evolution, arguing however that it is not necessary
to explain magnetar activity; our focus is different, as we suggest
plastic flow is \emph{inevitable} in some circumstances, and that its
effects are consistent with magnetar phenomena.

We suggest that plastic flow will dominate NS crustal field evolution for $B\gtrsim
10^{15}$ G, compete with Hall drift in the outer crust for $B\sim
10^{14}$ G, and probably be mostly irrelevant for $B\lesssim 10^{13}$ G. This suggests
that it plays a key role for young magnetars, in particular. Plastic
flow is an unusual field-evolution mechanism, since it 
shuts down below a certain (density-dependent) field
strength -- meaning that there may be a genuine distinction in the
underlying physics of some magnetars compared with other NSs, and
presenting an additional challenge to attempts (e.g. \citet{vigano}) to `unify' the
different observational manifestations of NSs.


\acknowledgements

I thank Ian Jones and Danai Antonopoulou for useful discussions, Nils Andersson and Wynn
Ho for comments on a draft of this paper, and the anonymous referee
for a number of helpful suggestions to improve the presentation of
this work. I acknowledge support from STFC via grant number ST/M000931/1.


\end{document}